\documentclass[aps,prl,twocolumn,floatfix,footinbib,showpacs]{revtex4}
\usepackage{graphicx}
\usepackage{amsfonts,amsmath,amssymb}
\usepackage{amsthm}
\usepackage{dsfont,bm}
\usepackage{color}
\usepackage{soul} 
\usepackage{amsbsy}
\usepackage[colorlinks=true,linkcolor=blue,pagecolor=blue,filecolor=blue,menucolor=blue,urlcolor=blue,citecolor=blue,anchorcolor=blue]{hyperref}%
\usepackage{sidecap}

\begin{document}

\title{Nonlocal Andreev Entanglements and Triplet Correlations in Graphene with Spin Orbit Coupling}

\author{Razieh Beiranvand }
\affiliation{Department of Physics, K.N. Toosi University of Technology, Tehran 15875-4416, Iran}
\author{Hossein Hamzehpour}
\affiliation{Department of Physics, K.N. Toosi University of Technology, Tehran 15875-4416, Iran}
\author{Mohammad Alidoust}
\affiliation{Department of Physics, K.N. Toosi University of Technology, Tehran 15875-4416, Iran}
\date{\today} 
\begin{abstract}
Using a wavefunction Dirac Bogoliubov-de Gennes method, we demonstrate that the tunable Fermi level of a graphene layer in the presence of Rashba spin orbit coupling (RSOC) allows for producing an anomalous nonlocal Andreev reflection and equal spin superconducting triplet pairing. We consider a graphene junction of a ferromagnet-RSOC-superconductor-ferromagnet configuration and study scattering processes, the appearance of spin triplet correlations, and charge conductance in this structure. We show that the anomalous crossed Andreev reflection is linked to the equal spin triplet pairing. Moreover, by calculating current cross-correlations, our results reveal that this phenomenon causes negative charge conductance at weak voltages and can be revealed in a spectroscopy experiment, and may provide a tool for detecting the entanglement of the equal spin superconducting pair correlations in hybrid structures.
\end{abstract}
\maketitle

\textit{Introduction}- Superconductivity and its hybrid structures with other phases can host a wide variety of intriguing fundamental phenomena and functional applications such as Higgs mechanism ~\cite{Pekker2015ARCM}, Majorana fermions~\cite{Beenakker2013ARCM}, topological quantum computation~\cite{Nayak2008RMP}, spintronics~\cite{Eschrig2015RPP}, and quantum entanglement~\cite{qc,blatter,raimond,aspect}. 
The quantum entanglement describes quantum states of correlated objects with nonzero distances \cite{qc,raimond} that are expected to be employed in novel ultra-fast technologies such as secure quantum computing \cite{qc,Nayak2008RMP}.

From the perspective of BCS theory, $s$-wave singlet superconductivity is a bosonic phase created by the coupling of two charged particles with opposite spins and momenta (forming a so-called Cooper pair) through an attractive potential \cite{bcs}. The two particles forming a Cooper pair can spatially have a distance equal or less than a coherence length $\xi_S$ \cite{bcs}. Therefore, a Cooper pair in the BCS scenario can serve as a natural source of entanglement with entangled spin and momentum. As a consequence, one can imagine a heterostructure made of a single $s$-wave superconductor and multiple nonsuperconducting electrodes in which an electron and hole excitation from different electrodes are coupled by means of a nonlocal Andreev process \cite{blatter,feinberg,recher,cpr_epx1,Byers1995PRL}. This idea has so far motivated numerous theoretical and experimental endeavours to explore this entangled state in various geometries and materials \cite{Chtchelkatchev,Kalenkov,cpr_epx2,cpr_epx3,cpr_epx4,cpr_epx5,Amitai,Walter,Hussein,Cayssol2008PRL,Linder2008PRB,Bignon2004EL,recher,moghadam,saha,lee}. Nonetheless, the nonlocal Andreev process is accompanied by an elastic cotunneling current that makes it practically difficult to detect unambiguously the signatures of nonlocal entangled state \cite{cpr_epx2,cpr_epx3,cpr_epx4,cpr_epx5,feinberg,cpr_epx1,Byers1995PRL}. This issue however may be eliminated by making use of a graphene based hybrid device that allows for locally controlled Fermi level \cite{Cayssol2008PRL}.

On the other hand, the interplay of $s$-wave superconductivity and an inhomogeneous magnetization can convert the superconducting spin singlet correlations into equal spin triplets \cite{Buzdin2005,first}. After the theoretical prediction of the spin triplet superconducting correlations much effort has been made to confirm their existence \cite{Eschrig2015RPP,linder, halt2,Moor,Khaydukov,Halterman2007,Keizer2006,crnt_1,crnt_2,valve1,Baker,gol1,bergeret1,alidoust1,alidoust2}. For example, a finite supercurrent was observed in a half-metallic junction that was attributed to the generation of equal spin triplet correlations near the superconductor-half metal interface \cite{Keizer2006}. Also, it was observed that in a Josephson junction made of a Holmium-Cobalt-Holmium stack, the supercurrent as a function of the Cobalt layer decays exponentially without any sign reversals due to the presence of equal spin triplet pairings \cite{crnt_1,crnt_2}.
One more signature of the equal spin triplet pairings generated in the hybrid structures may be detected in superconducting critical temperature \cite{singh,half,bernard1,gol1} and density of states \cite{bernard2,zep,zep2,zep3}. Nevertheless, a direct observation of the equal spin triplet pairings in the hybrid structures is still lacking.

In this paper, we show that the existence of the equal spin superconducting triplet correlations can be revealed through charge conductance spectroscopy of a graphene based ferromagnet - Rashba SOC - superconductor - ferromagnet junction. We study all possible electron/hole reflections and transmissions in such a configuration and show that by tuning the fermi level a regime is accessible in which spin reversed cotunneling and usual crossed Andreev reflections are blocked while a conventional cotunneling and anomalous nonlocal Andreev channel is allowed. We justify our findings by analyzing the bandstructure of the system. Moreover, we calculate various superconducting correlations and show that, in this regime, the equal spin triplet correlation has a finite amplitude while the unequal spin triplet component vanishes. Our results show that the anomalous crossed Andreev reflection results in a negative charge conductance at low voltages applied across the junction and can be interpreted as evidence for the generation and entanglement of equal spin superconducting triplet correlations in hybrid structures \cite{equalspin1,equalspin2,equalspin3,equalspin4,equalspin5}.        

\begin{figure}
\includegraphics[width=8.5cm, height=3.20 cm]{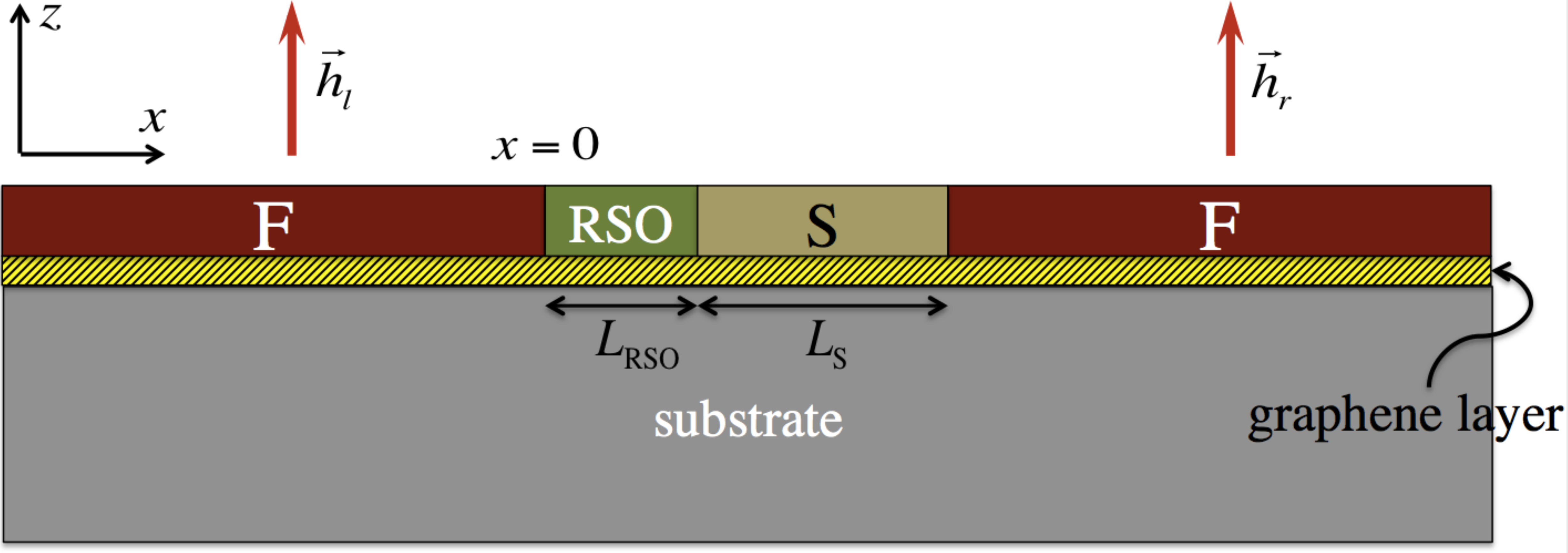}
\caption{Schematic of the graphene based F-RSO-S-F hybrid. The system resides in the $xy$ plane and the junctions are located along the $x$ axis. The length of the RSO and S regions are denoted by $L_{\text{RSO}}$ and $L_{\text{S}}$. The magnetization of the F regions ($\vec{h}_{l,r}$) are assumed fixed along the $z$ axis. We assume that the ferromagnetism, spin orbit coupling, and superconductivity is induced into the graphene layer by means of the proximity effect.}
\label{fig1}
\end{figure}

\textit{Method and Results}- As seen in Fig. \ref{fig1}, we assume that the ferromagnetism, superconductivity, and spin orbit coupling are separately induced into the graphene layer through the proximity effect as reported experimentally in Refs. \onlinecite{ex1,avsar2014nat,Heersche2007Nature} for isolated samples. Therefore, the low energy behaviour of quasiparticles, quantum transport characteristics, and thermodynamics of such a system can be described by the Dirac Bogoliubov-de Gennes (DBdG) formalism \cite{beenakkerrmp,halt2}:
	\begin{equation}
	\fontsize{8}{2}
	\left(\begin{array}{cc}
	\mathcal{H}_\text{D}+\mathcal{H}_i-\mu^i & \Delta e^{i\phi}\\
	\Delta^{*}e^{-i\phi} & \mu^i-{\cal T}[\mathcal{H}_\text{D}-\mathcal{H}_i]{\cal T}^{-1}\\
	\end{array}\right)
	\left(\begin{array}{c}
	u \\
	v \\
	\end{array}\right)=\varepsilon
	\left(\begin{array}{c}
	u \\
	v \\
	\end{array}\right),
	\label{Eq.DBdG}
	\end{equation}
in which $\varepsilon$ is the quasiparticles' energy and $\mathcal{T}$ represents a time-reversal operator\cite{beenakkerrmp,halt2}. Here $\mathcal{H}_D=\hbar v_F s_0 \otimes(\sigma_x k_x+\sigma_y k_y)$ with $v_F$ being the fermi velocity \cite{beenakkerrmp}. $s_{x,y,z}$ and $\sigma_{x,y,z}$ are 2$\times$2 Pauli matrices, acting on the spin and pseudo-spin degrees of freedom, respectively. The superconductor region with a macrospic phase $\phi$ is described by a gap $\Delta$ in the energy spectrum. The chemical potential in a region $i$ is shown by $\mu^i$ while the corresponding Hamiltonians read:
	\begin{equation}
	\fontsize{8}{2}
	\mathcal{H}_i=
	\begin{cases}
	\mathcal{H}_{\text{F}}=h_l(s_z\otimes\sigma_0) & x\leq 0 \\
	\mathcal{H}_{\text{RSO}}=\lambda (s_y\otimes \sigma_x-s_x \otimes \sigma_y) & 0 \leq x \leq L_{\text{RSO}} \\
	\mathcal{H}_{\text{S}}=-U_0 (s_0 \otimes \sigma_0) & L_{\text{RSO}}\leq x \leq L_\text{S}+L_{\text{RSO}} \\
	\mathcal{H}_{\text{F}}=h_r(s_z \otimes \sigma_0) & L_\text{S}+L_{\text{RSO}} \leq x
	\end{cases}.
	\label{Eq.Hamiltonians}
	\end{equation}
The magnetization $\vec{h}_{l,r}$ in the ferromagnet segments are assumed fixed along the $z$ direction with a finite intensity $h_{l,r}$. $\lambda$ is the strength of Rashba spin orbit coupling and $U_0$ is an electrostatic potential in the superconducting region. Previous self-consistent calculations have demonstrated that sharp interfaces between the regions can be an appropriate approximation \cite{halt2,halt1,beenakkerrmp,efetov,Shalom} 
The length of the RSO and S regions are $L_{\text{RSO}}$ and $L_{\text{S}}$, respectively.

To determine the properties of the system, we diagonalize the DBdG Hamiltonian Eq. (\ref{Eq.DBdG}) in each region and obtain corresponding eigenvalues:
	\begin{equation}
	\fontsize{7}{2}
	\varepsilon=\begin{cases}
	\pm \mu^{\text{F}_l}\pm\sqrt{(k_x^{\text{F}_l})^2+q_n^2}\pm h_l & x\leq 0\\
	\pm \mu^{\text{RSO}}\pm\sqrt{(k_x^{\text{RSO}})^2+q_n^2+\lambda^2}\pm \lambda & 0 \leq x \leq L_{\text{RSO}} \\
	\pm \sqrt{(\mu^S+U_0\pm\sqrt{(k_x^S)^2+q_n^2})^2+|\Delta_0|^2}& L_{\text{RSO}} \leq x \leq L_{\text{RSO}}+L_\text{S} \\
	\pm \mu^{\text{F}_r}\pm\sqrt{(k_x^{\text{F}_r})^2+q_n^2}\pm h_r & L_{\text{RSO}}+L_\text{S}\leq x
	\end{cases}.
	\label{Eq.Eigenvalues}
	\end{equation}
The associated eigenfunctions are given in appendix. The wavevector of a quasiparticle in region $i$ is $\textbf{k}_i=(k_x^i,q_n)$ so that its transverse component is assumed conserved upon scattering. In what follows, we consider a heavily doped superconductor $U_0\gg \varepsilon, \Delta$ which is an experimentally relevant regime \cite{beenakkerrmp}. We also normalize energies by the superconducting gap at zero temperature $\Delta_0$ and lengths by the superconducting coherent length $\xi_S=\hbar v_F/\Delta_0$.

\begin{figure}
	\includegraphics[width=8.8cm, height=6.0 cm
	]{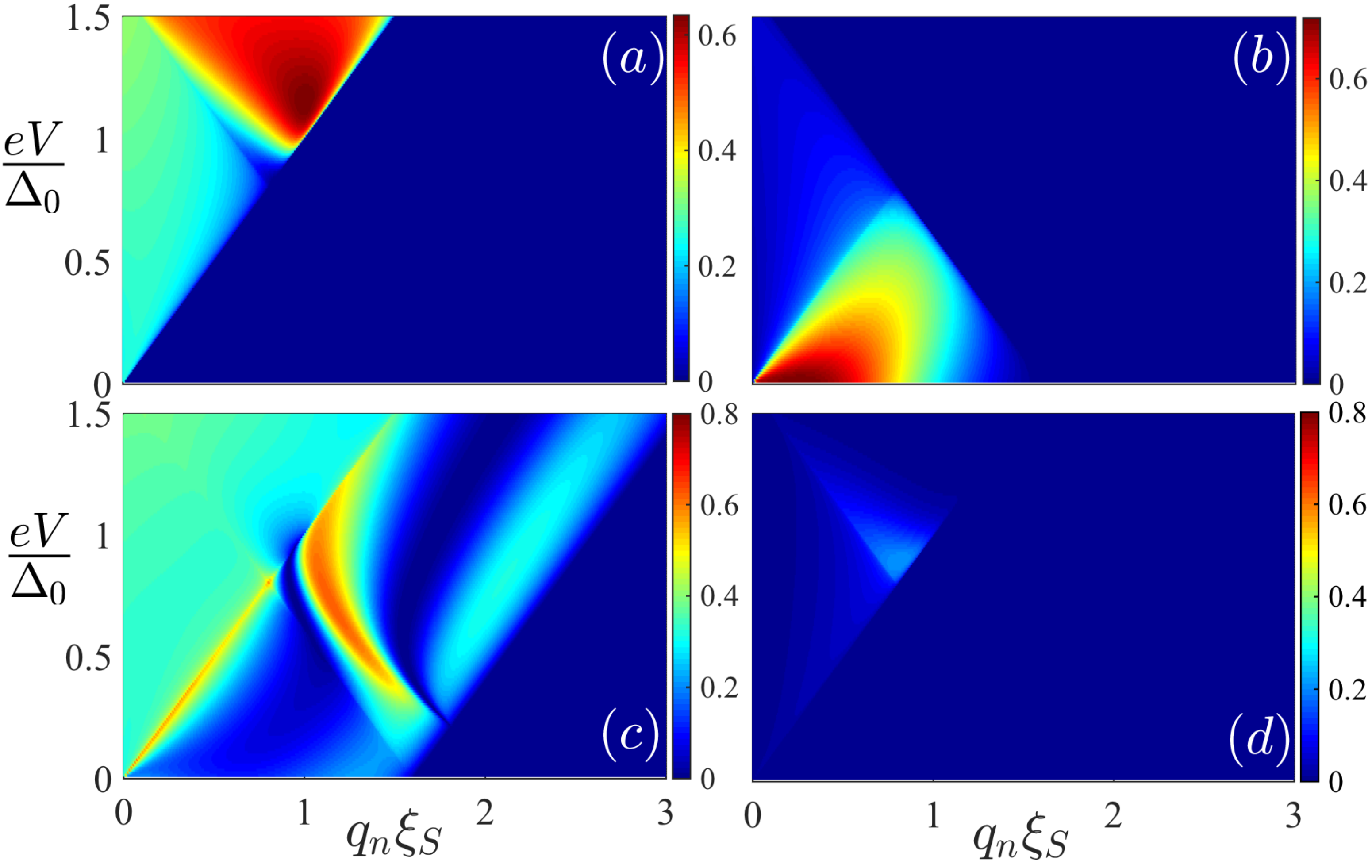}
	\caption{ ($a$) Spin-reversed cotunneling probability $|t_{e}^{\downarrow}|^2$, ($b$) Anomalous crossed Andreev reflection probability $|t_{h}^{\uparrow}|^2$, ($c$) Conventional cotunneling $|t_{e}^{\uparrow}|^2$, ($d$) Usual CAR $|t_{h}^{\downarrow}|^2$. The probabilities are plotted vs the transverse component of wavevector $q_n$ and voltage bias across the junction $eV$. We set $\mu^{\text{F}_l}=\mu^{\text{F}_r}=h_l=h_r=0.8\Delta_0, \mu^{\text{RSO}}=2.6\Delta_0, \lambda=\Delta_0, L_{\text{RSO}}=0.5\xi_S, L_{\text{S}}=0.4\xi_S$.}
	\label{fig2}
	\end{figure}

Since the magnetization in F regions is directed along the $z$ axis, which is the quantization axis, it allows for unambiguously analyzing spin-dependent processes. Therefore, we consider a situation where an electron with spin-up (described by wavefunction $\psi_{e,\uparrow}^{\text{F},+}$) hits the RSO interface at $x=0$ due to a voltage bias applied. This particle can reflect back ($\psi_{e,\uparrow (\downarrow)}^{\text{F},-}$) with probability amplitude $r_N^{\uparrow(\downarrow)}$ or enter the superconductor as a Cooper pair and a hole ($\psi_{h,\uparrow (\downarrow)}^{\text{F},-}$) with probability amplitude $r_A^{\uparrow(\downarrow)}$ reflects back which is the so called Andreev reflection. Hence, the total wavefunction in the left F region is (see appendix and Ref. \onlinecite{equalspin3}):
	\begin{equation}
	\begin{array}{cc}
	\mathbf{\Psi}^{\text{F}_l}(x)= & \psi_{e,\uparrow}^{\text{F},+}(x)+r_{N}^\uparrow\psi_{e,\uparrow}^{\text{F},-}(x)
	+ r_{N}^{\downarrow}\psi_{e,\downarrow}^{\text{F},-}(x)\\ &+r_{A}^{\downarrow}\psi_{h,\downarrow}^{\text{F},-}(x)
	+r_{A}^{\uparrow}\psi_{h,\uparrow}^{\text{F},-}(x).
	\end{array}
	\label{Eq.WFinFL}
	\end{equation}
The total wavefunction in the RSO and S parts are superpositions of right and left moving spinors with different quantum states $n$; $\psi^{\text{RSO}}_{n}$ and $\psi^{\text{S}}_{n}$ (see appendix): $\mathbf{\Psi}^{\text{RSO}}(x)= \sum_{n=1}^{8} a_n\psi^{\text{RSO}}_{n}(x)$ and $\mathbf{\Psi}^{\text{S}}(x)= \sum_{n=1}^{8} b_n\psi^{\text{S}}_{n}(x)$, respectively. The incident particle eventually can transmit into the right F region as an electron or hole ($\psi_{e,\uparrow\downarrow}^{\text{F},+}, \psi_{h,\uparrow\downarrow}^{\text{F},+}$) with probability amplitudes $t_e^{\uparrow\downarrow}$ and $t_h^{\uparrow\downarrow}$:
	\begin{equation}
	\mathbf{\Psi}^{\text{F}_r}(x)= t_{e}^{\uparrow}\psi_{e,\uparrow}^{\text{F},+}(x)+t_{e}^{\downarrow}\psi_{e,\downarrow}^{\text{F},+}(x)
	+ t_{h}^{\downarrow}\psi_{h,\downarrow}^{\text{F},+}(x) +t_{h}^{\uparrow}\psi_{h,\uparrow}^{\text{F},+}(x).
	\label{Eq.WFinFR}
	\end{equation}
The transmitted hole is the so called crossed Andreev reflection (CAR). By matching the wavefunctions at F-RSO, RSO-S, and S-F interfaces we obtain the probabilities described above. Figure \ref{fig2} exhibits the probabilities of usual electron cotunneling  $|t_e^\uparrow|^2$, spin-flipped electron $|t_e^\downarrow|^2$, usual crossed Andreev reflection  $|t_h^\downarrow|^2$, and anomalous crossed Andreev reflection $|t_h^\uparrow|^2$. To have a strong anomalous CAR signal, we set $L_\text{S}=0.4\xi_S$ which is smaller than the superconducting coherence length and $L_\text{RSO}=0.5\xi_S$ \cite{feinberg}. We also choose $\mu^{\text{F}_l}=\mu^{\text{F}_r}=h_l=h_r=0.8 \Delta_0, \mu^{\text{RSO}}=2.6\Delta_0, \lambda=\Delta_0$ and later clarify physical reasons behind this choice using band structure analyses. In terms of realistic numbers, if the superconductor is Nb \cite{Shalom} with a gap of the order of $\Delta_0\sim 1.03$meV and coherence length $\xi_S\sim 10$nm, the chemical potentials, magnetization strengths, and the RSO intensity are $\mu^{\text{F}_l}=\mu^{\text{F}_r}=h_l=h_r=0.824$meV, $\mu^{\text{RSO}}=2.68$meV, $\lambda=1.03$meV, respectively \cite{ex1,avsar2014nat}, and $L_\text{S}=4$nm, $L_\text{RSO}=5$nm. We see that the anomalous CAR has a finite amplitude and its maximum is well isolated from the other transmission channels in the parameter space. Therefore, by tuning the local Fermi levels the system can reside in a regime that allows for a strong signal of the anomalous CAR. According to Fig. \ref{fig2} this regime is accessible at low voltages $eV\ll \Delta_0$.      
The eigenvalues Eqs. (\ref{Eq.Eigenvalues}) determine the propagation critical angles of moving particles through the junction. By considering the conservation of transverse component of wavevector throughout the system, we obtain the following critical angles \cite{beenakkerrmp}:
\begin{subequations}
\begin{eqnarray}
\alpha_{e,\downarrow}^c=\arcsin\Big|\frac{\varepsilon+\mu^{F_r}-h_r}{\varepsilon+\mu^{F_l}+h_l}\Big|,\\\alpha_{h,\downarrow}^c=\arcsin\Big|\frac{\varepsilon-\mu^{F_r}+h_r}{\varepsilon+\mu^{F_l}+h_l}\Big|,\\\alpha_{e,\uparrow}^c=\arcsin\Big|\frac{\varepsilon+\mu^{F_r}+h_r}{\varepsilon+\mu^{F_l}+h_l}\Big|,\\\alpha_{h,\uparrow}^c=\arcsin\Big|\frac{\varepsilon-\mu^{F_r}-h_r}{\varepsilon+\mu^{F_l}+h_l}\Big|.
\end{eqnarray}\label{angles}
\end{subequations}
These critical angles are useful in calibrating the device properly for a regime of interest. For the spin-reversed cotunneling, the critical angle is denoted by $\alpha_{e,\downarrow}^c$ while for the conventional CAR we show this quantity by $\alpha_{h,\downarrow}^c$. Hence, to filter out these two tranmission channels, we set $\mu^{F_r}=h_r$ and choose a representative value $0.8\Delta_0$. In this regime, we see that $\alpha_{e(h),\downarrow}^c\rightarrow 0$ at low energies i.e. $\mu^{F_r}, h_r, \Delta \gg \varepsilon\rightarrow 0$ and thus, the corresponding transmissions are eliminated. This is clearly seen in Figs. \ref{fig2}(a) and \ref{fig2}(d) at $eV\ll\Delta_0$. At the same time, the critical angles to the propagation of conventional electron cotunneling and anomalous crossed Andreev reflection reach near their maximum values $\alpha_{e(h),\uparrow}^c\rightarrow \pi/2$ consistent with Figs. \ref{fig2}(b) and \ref{fig2}(c). We have analyzed the reflection and transmission processes using a bandstructure plot, presented in appendix, that can provide more sense on how a particle is scattered in this regime.  

	 \begin{figure}
	 \includegraphics[width=7cm, height=6.0 cm]{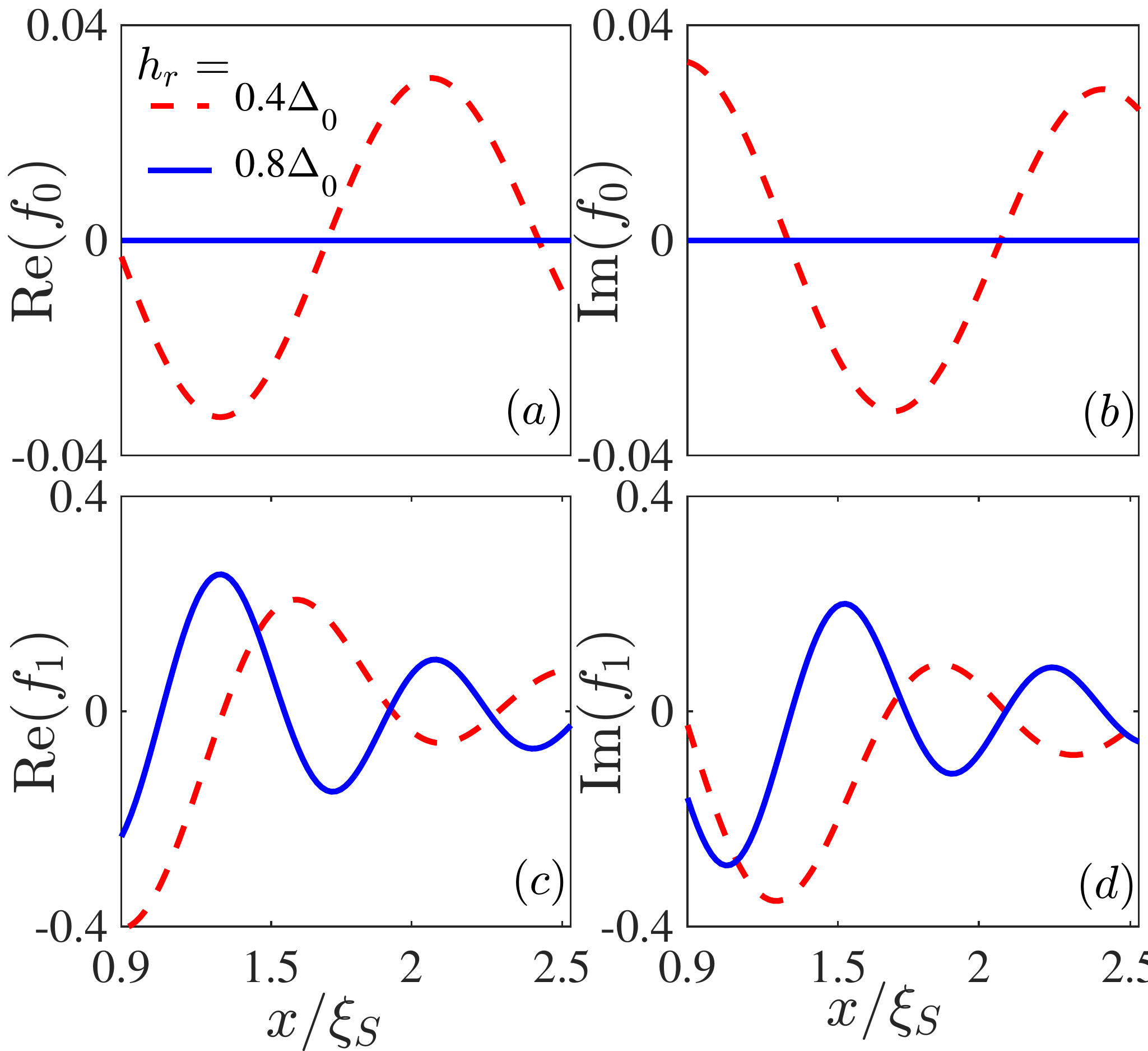}
	 \caption{ ($a$)-($d$) real and imaginary parts of opposite spin $f_0$ and equal spin pairings $f_1$ within the $F_r$ region $x \geq L_{\text{RSO}}+L_{\text{S}}$ at weak voltages $eV\ll\Delta_0$. The parameter values are the same as those of Fig. \ref{fig2} except we now compare two cases where $\mu^{F_l}=\mu^{F_r}=h_l=0.8\Delta_0$ and $h_r=0.4\Delta_0, 0.8\Delta_0$.}
	 \label{fig3}
	 \end{figure}

To gain better insights into the anomalous CAR, we calculate the opposite ($f_0$) and equal ($f_1$) spin pair correlations in the $F_r$ region \cite{Halterman2007,halt2}:
\begin{subequations}
\begin{eqnarray}
f_0(x,t)= && +\frac{1}{2}\sum\limits_\beta \xi(t)[u_{\beta, K}^\uparrow v_{\beta,K'}^{\downarrow,*}+u_{\beta,K'}^{\uparrow}v_{\beta,K}^{\downarrow *}\nonumber\\
&&- u_{\beta,K}^\downarrow v_{\beta,K'}^{\uparrow *}-u_{\beta,K'}^\downarrow v_{\beta,K}^{\uparrow *}],\\
f_1(x,t)= && -\frac{1}{2}\sum\limits_\beta \xi(t)[u_{\beta, K}^\uparrow v_{\beta,K'}^{\uparrow,*}+u_{\beta,K'}^{\uparrow}v_{\beta,K}^{\uparrow *}\nonumber\\
&&+ u_{\beta,K}^\downarrow v_{\beta,K'}^{\downarrow *}+u_{\beta,K'}^\downarrow v_{\beta,K}^{\downarrow *}],
\end{eqnarray}\label{Eq.Triplet01}
\end{subequations}
where $K$ and $K'$ denote different valleys and $\beta$ stands for A and B sub-lattices \cite{halt2,beenakkerrmp}. Here, $\xi(t)=\cos(\varepsilon t)-i \sin(\varepsilon t)\tanh(\varepsilon/2T)$ and $t$ is the relative time in the Heisenberg picture and $T$ is the temperature of the system \cite{halt2,Halterman2007}. Figure \ref{fig3} shows the real and imaginary parts of opposite and equal spin pairings in the $F_r$ region, extended from $x=L_\text{RSO}+L_\text{S}$ to infinity, at $eV\ll\Delta_0$. For the set of parameters corresponding to Fig. \ref{fig2}, we see that $f_0$ pair correlation is vanishingly small, while the equal spin triplet pair correlation $f_1$ has a finite amplitude. We also plot these correlations for a different set of parameters where $\mu^{F_l}=\mu^{F_r}=h_l=0.8\Delta_0$, while $h_r=0.4\Delta_0$. The opposite spin triplet pairing $f_0$ is now nonzero too. Therefore, at low voltages and the parameter set of Fig. \ref{fig2}, the nonvanishing triplet correlation is $f_1$, which demonstrates the direct link of $f_1$ and $t_h^{\uparrow}$. This direct connection can be proven by looking at the total wavefunction in the right F region, Eq. (\ref{Eq.WFinFR}), transmission probabilities shown in Fig. \ref{fig2}, and the definition of triplet correlations, Eqs. (\ref{Eq.Triplet01}). One can show that when $t^\downarrow_e$ and $t^\downarrow_h$ vanish, $f_0$ disappears and $f_1$ remains nonzero, which offers a spin triplet valve effect.   

\begin{figure}
\includegraphics[scale=0.38]{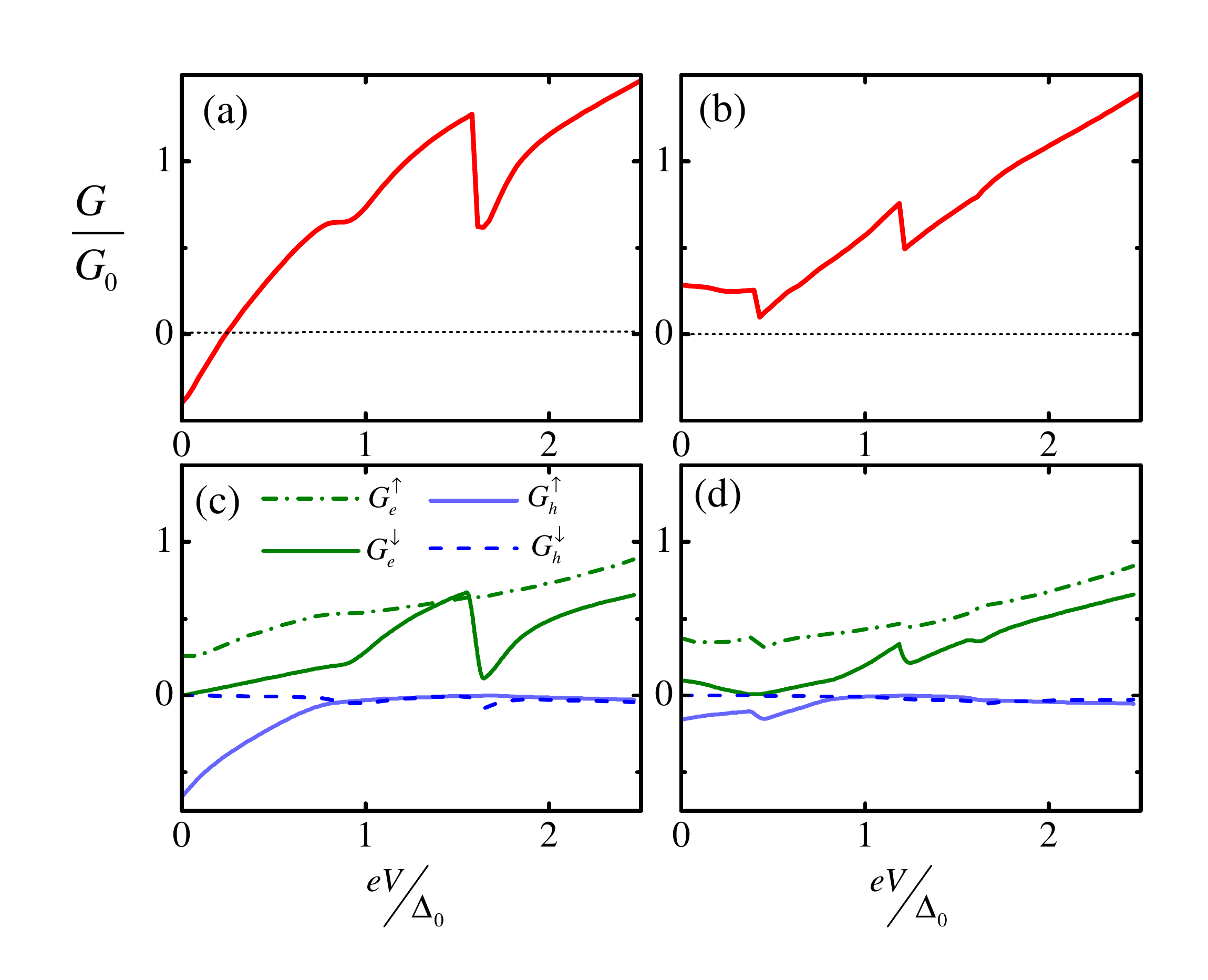}
\caption{ Charge conductance (top panels) and its components (bottom panels). ($a$) and ($c$) charge conductance associated with the probabilities presented in Figs. \ref{fig2} and \ref{fig3} ($h_r=0.8\Delta_0$) and its components, respectively. ($b$) and ($d$) the same as panels ($a$) and ($c$) except we now consider $h_r=0.4$ (see Fig. \ref{fig3}). The conductance is normalized by $G_0=G_\uparrow+G_\downarrow$.}
\label{fig4}
\end{figure}

We calculate the charge conductance through the BTK formalism:
\begin{equation}
G=\int dq_n\sum_{s=\uparrow,\downarrow} G_s \Big(|t_{e}^{s}|^2-|t_{h}^{s}|^2\Big),
\label{Eq.Conductance}
\end{equation}
where we define $G_{\uparrow\downarrow}=2 e^2|\varepsilon+\mu_l\pm h_l|W/h\pi$ in which $W$ is the width of the junction. Figures \ref{fig4}(a) and \ref{fig4}(b) exhibit the charge conductance as a function of bias voltage $eV$ across the junction at $h_r=0.8\Delta_0$ and $0.4\Delta_0$, while the other parameters are set the same as those of Figs. \ref{fig2} and \ref{fig3}. As seen, the charge conductance is negative at low voltages when $h_r=0.8\Delta_0$, whereas this quantity becomes positive for $h_r=0.4\Delta_0$. To gain better insights, we separate the charge conductance into $G_{e,(h)}^{\uparrow\downarrow(\uparrow\downarrow)}$, corresponding to the transmission coefficients $t_{e,(h)}^{\uparrow\downarrow(\uparrow\downarrow)}$ used in Eq. (\ref{Eq.Conductance}). Figures \ref{fig4}(c) and \ref{fig4}(d) illustrate the contribution of different transmission coefficients into the conductance. We see in Fig. \ref{fig4}(c) that $G_h^{\uparrow}$ dominates the other components and makes the conductance negative. As discussed earlier, this component corresponds to the anomalous CAR which is linked to the equal spin triplet pairing, Fig. \ref{fig3}. This component however suppresses when $h_r=0.4\Delta_0$ so that the other contributions dominate, and therefore the conductance is positive for all energies. Hence, the nonlocal anomalous Andreev reflection found in this work can be revealed in a charge conductance spectroscopy. There are also abrupt changes in the conductance curves that can be fully understood by analyzing the band structure. We present such an analysis in appendix.

In line with the theoretical works summarized in Ref. \onlinecite{beenakkerrmp}, we have neglected spin-dependent and -independent impurities and disorders as well as substrate and interface effects in our calculations \cite{impur1,impur2,impur3}. Nonetheless, a recent experiment has shown that such a regime is accessible with today's equipments \cite{Shalom}. Moreover, the same assumptions have already resulted in fundamentally important predictions such as the Specular Andreev reflection \cite{beenakkerrmp} that recently was observed in experiment \cite{efetov}. The experimentally measured mean free path of moving particles in a monolayer graphene deposited on top of a hexagonal boron nitride substrate is around $\ell\sim 140$nm\cite{Bretheau}. The coherence length of induced superconductivity into a monolayer graphene using a Nb superconductor was reported as $\xi_S\sim 10$nm \cite{Shalom}. In this situation, where $\ell\gg\xi_S$, the Andreev mechanism is experimentally relevant. On the other hand, it has been demonstrated that the equal-spin pairings discussed here are long-range and can survive even in systems with numerous strong spin-independent scattering resources \cite{bergeret1,alidoust1,alidoust2}. Therefore, as far as the Andreev mechanism is a relevant scenario in a graphene-based F-RSO-S-F device containing spin-independent scattering resources, i.e. $\ell\gg\xi_S$, we expect that the negative conductance explored in this paper is experimentally accessible.

In conclusion, motivated by recent experimental achievements in the induction of spin orbit coupling into a graphene layer \cite{avsar2014nat,ex1}, we have theoretically studied quantum transport properties of a graphene based ferromagnet-RSOC-superconductor-ferromagnet junction. Our results reveal that by manipulating the Fermi level in each segment, one can create a dominated anomalous crossed Andreev reflection. We calculate the charge conductance of the system in this regime and show that this phenomenon results in negative charge conductance at low voltages. By calculating various pairing correlations, we demonstrate a direct link between the appearance of anomalous CAR and equal spin triplet correlations. Our findings suggest that a conductance spectroscopy of such a junction can detect the signatures of the anomalous CAR and entanglement of equal spin superconducting triplet pairings in hybrid structures.

\textit{Acknowledgments}- We are grateful to M. Salehi for valuable and helpful discussions. M.A. also thanks K. Halterman for useful conversations.

\begin{widetext}

\section{APPENDIX}

\section{Wavefunctions and critical angles}

The eigenfunctions in the F regions can be expressed by:
  \begin{equation}
 \begin{array}{l}
\psi^{\text{F},\pm}_{e,\uparrow}(x)=\left(\mathbf{0^2}, 1 , \pm e^{\pm i
    \alpha_\uparrow^e}, \mathbf{0^4}\right)^{\bf T} e^{\pm i
  k_{x,\uparrow}^{\text{F},e} x},\;\;\;\;\;
\psi^{\text{F},\pm}_{e,\downarrow}(x)=\left( 1 , \pm e^{\pm i \alpha_\downarrow^e}, \mathbf{0^2},\mathbf{0^4}\right)^{\bf T} e^{\pm i k_{x,\downarrow}^{\text{F},e} x},\\
\\
   \psi^{\text{F},\pm}_{h,\uparrow}(x)= \left(\mathbf{0^4}, 1 ,  \mp
     e^{ \pm i \alpha_\uparrow^h},\mathbf{0^2}\right)^{\bf T} e^{ \pm
     i k_{x,\uparrow}^{\text{F},h} x},\;\;\;\;\;
 \psi^{\text{F},\pm}_{h,\downarrow}(x)=\left( \mathbf{0^4},
   \mathbf{0^2}, 1 , \mp e^{ \pm i\alpha_\downarrow^h}\right)^{\bf T}
 e^{\pm i k_{x,\downarrow}^{\text{F},h} x},
  \end{array}
     \end{equation} 
where $\mathbf{0^n}$ is a $1 \times n$ zero matrix and ${\bf T}$ is a transpose operator. The
junction width $W$ is assumed wide enough so that the $y$ component of wavevector $k_y\equiv q_n$ is conserved and we factor out the corresponding
multiplication i.e. $\exp(i q_n y)$. 
The $\alpha_{\uparrow, (\downarrow)}^{e(h)}=\arctan ({q_n}/{ k_{x, \uparrow(\downarrow)}^{\text{F},e(h)}} )$ variables are the
propagation angles for electron and hole excitations with a given spin orientation. 
The $x$ component of wavevectors are not conserved upon
scattering and is given by $\hbar v_F k_{x,{\uparrow\downarrow}}^{\text{F},e}=(\varepsilon+\mu^\text{F}\pm h)\cos\alpha_{\uparrow\downarrow}^e$ and $\hbar v_F k_{x,{\uparrow\downarrow}}^{\text{F},h}=(\varepsilon-\mu^\text{F}\mp h)\cos\alpha_{\uparrow\downarrow}^h$. The wavefunctions are not propagating for larger values of $q_n$ than a critical value $q^c$.
The critical values for electrons and holes with spin-up/-down are $q_{e,\uparrow\downarrow}^c=|\varepsilon+\mu^\text{F}\pm h|/\hbar v_F$ and $q_{h,\uparrow\downarrow}^c=|\varepsilon-\mu^\text{F}\mp h|/\hbar v_F$.

\*
\*
\\
The energy spectrum in
the RSO region is gapless with a splitting of magnitude
$2\lambda$ between its subbands. 
The eigenfunctions in the RSO part are: 
  \begin{equation}
  \begin{array}{c}
  \psi^{\text{RSO},\pm}_{e,\eta=+1}(x)= \Big (\mp i f_{+}^e e^{\mp i
      \theta_{+}^e}, -i, 1, \pm f_{+}^e e^{\pm i
      \theta_{+}^e},\mathbf{0^4}\Big)^{\bf T}e^{\pm i k_{x,+}^{\text{RSO},e}x},\\
  \psi^{\text{RSO},\pm}_{e, \eta=-1}(x)=\Big(\pm f_{-}^e e^{\mp i \theta_{-}^e}, 1, -i, \mp if_{-}^e e^{\pm i \theta_{-}^e},\mathbf{0^4}\Big)^{\bf T}e^{\pm i k_{x,-}^{\text{RSO},e}x}\\
 \psi^{\text{RSO},\pm}_{h,\eta=+1}(x)= \Big (\mathbf{0^4},\mp i f_{+}^h e^{\mp i \theta_{+}^h}, -i, 1, \pm f_{+}^h e^{\pm i \theta_{+}^h}\Big)^{\bf T}e^{\pm i k_{x,+}^{\text{RSO},h}x},\\
  \psi^{\text{RSO}, \pm}_{h, \eta=-1}(x)=\Big (\mathbf{0^4}, \pm f_{-}^h e^{\mp i \theta_{-}^h}, 1, -i, \mp if_{-}^h e^{\pm i \theta_{-}^h}\Big)^{\bf T}e^{\pm i k_{x,-}^{\text{RSO},h}x}
  \end{array}.
  \label{Eq.13}
  \end{equation}
the $x$ component of wavevectors are given by $\hbar v_Fk_{x,\eta}^{\text{RSO},e(h)}=(\mu^\text{RSO}+(-)\varepsilon)f_{\eta}^{e(h)}\cos\theta_\eta^{e(h)}$ in which $f_\eta^{e(h)}=\sqrt{1+2\eta\lambda(\mu^\text{RSO}+(-)\varepsilon)^{-1}}$ and $\theta_{\eta}^{e(h)}=\arctan({q_n 
        }/{ k_{x, \eta}^{\text{RSO},e(h)}})$. Here $\theta_{\eta}^{e(h)}$ are the electron and hole propagation
angles in the RSO region.

\*
\* 
\\
The wavefunctions in the superconductor part within the heavily dopped
regime are given by:
   \begin{equation}
   \begin{array}{l}
   \psi^{\text{S},\pm}_{e, 1}(x)=\Big(e^{+i \beta}, \pm e^{+i
     \beta}, \mathbf{0^2}, e^{-i\phi}, \pm e^{-i\phi},
   \mathbf{0^2}\Big)^{\bf T}e^{\pm i k_x^{\text{S},e} x},\;\;\;\;
   \psi^{\text{S},\pm}_{e,2}(x)=\Big (\mathbf{0^2}, e^{+i \beta}, \pm e^{+i \beta}, \mathbf{0^2}, e^{-i\phi}, \pm e^{-i\phi}\Big)^{\bf T}e^{\pm i k_x^{\text{S},e} x},\\
   \\
   \psi^{\text{S},\pm}_{h, 1}(x)= \Big (e^{-i \beta}, \mp e^{-i \beta}, \mathbf{0^2}, e^{-i\phi}, \mp e^{-i\phi}, \mathbf{0^2}\Big)^{\bf T}e^{\pm i k_x^{\text{S},h} x},\;\;\;\;
   \psi^{\text{S}, \pm}_{h, 2}(x)= \Big (\mathbf{0^2}, e^{-i \beta}, \mp e^{-i \beta}, \mathbf{0^2}, e^{-i\phi}, \mp e^{-i\phi}\Big)^{\bf T}e^{\pm i k_x^{\text{S},h} x}
   \end{array}
   \label{WF.S}.
   \end{equation} 
in which
    \begin{equation}
    \beta=\begin{cases}
    +~\arccos(\varepsilon/\Delta)&       \varepsilon\leq \Delta \\
    -i ~\text{arccosh}(\varepsilon/\Delta) &   \varepsilon \geq\Delta
    \end{cases}
    \label{Eq.3},
    \end{equation}
and the macroscopic phase of superconductor is shown by
$\phi$. Nonetheless, the phase of a single superconductor plays no
role in the quantum transport and we therefore set it zero in our
calculations.

\*
\* 
\\

\section{bandstructure discussions}

In order to determine the origines of the abrupt changes and
nonmonotonic behaviors seen in the charge conductance of Fig. \ref{fig4} in the
main text, we start with a simplified set of parameters where $\mu^{F_r}=\mu^{F_l}=0$,
$h_l=0$, and $h_r=0.8\Delta_0$. The other parameters are set identical
to those of used in Fig. \ref{fig4} i.e. $\mu^{\text{RSO}}=2.6\Delta_0$,
$\lambda=\Delta_0$, $L_{\text{RSO}}=0.5\xi_S$, and
$L_{\text{S}}=0.4\xi_S$. The charge conductance in this regime is shown in
Fig. \ref{fig1app}(a) and its constituting components are plotted in Fig. \ref{fig1app}(b). The
charge conductance is symmetric with respect to $eV=0$ and therefore we restrict our plots to $eV>0$. We see that the charge conductance shows two abrupt
changes at $eV=0.4\Delta_0$ and $0.8\Delta_0$. As seen from Fig. \ref{fig1app}(b),
the first abrupt change is caused by $G^\downarrow_\text{e}$ compoenet
while the second
abrupt change at $eV=0.8\Delta_0$ is originated from nonzero
components at this voltage bias i.e. $G^\uparrow_\text{e}$ and
$G^\downarrow_\text{h}$. To illustrate the physical origines and
corresponding processes resulting in these
abrupt changes we present the bandstructure analyses of the system
in Fig. \ref{fig2app}. The
low-energy bandstructure of each region is plotted and particles/holes
are shown by solid circles/circles. The inclusion of the particles/holes
symbols in Figs. \ref{fig2app} and \ref{fig3app} helps to easily follow scattering processes during the quantum
transport across the F-RSO-S-F junction. The horizontal solid line
shows the Fermi level and since we set $h_l=0$, the spin-up and -down
subbands are degenerated in the left region. An incident particle with
spin-up and
energy $\varepsilon$ (marked by A) can
reflect back as a particle with spin-up (B) or -down (D) or hole with
spin-up (C)
or -down (E). 
	\begin{figure*}
	\includegraphics[width=14.0cm, height=4.0 cm
	]{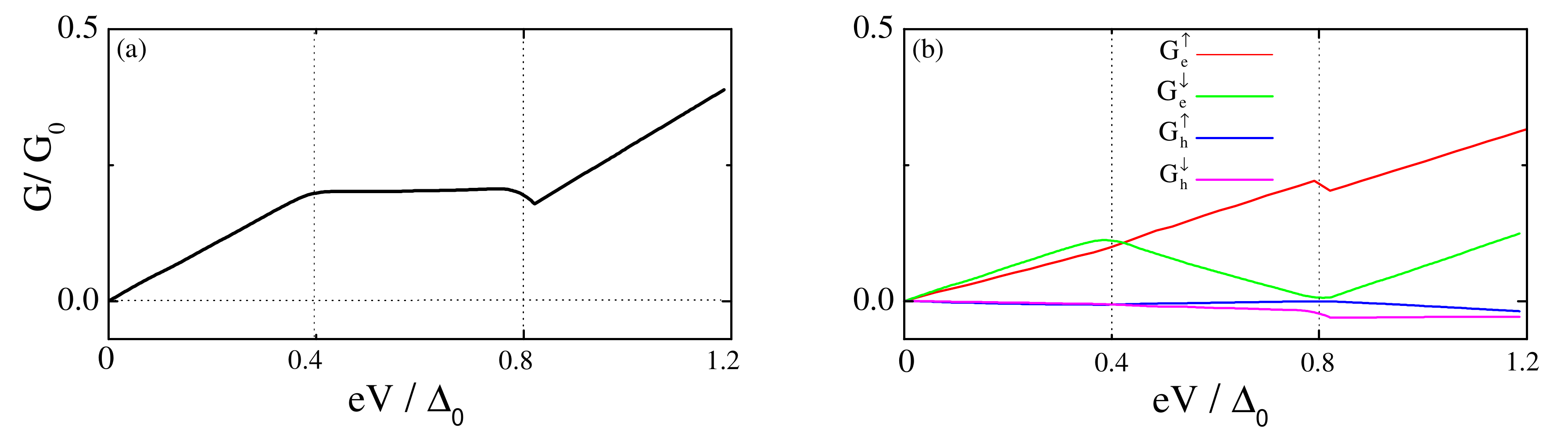}
	\caption{Charge conductance $G/G_0$ (a) and its constituting
          components
          (b), the same as Fig. \ref{fig4} in the main text, as a
          function of voltage bias $eV$ applied across the junction. Here, we
          set $\mu^{F_r}=\mu^{F_l}=0$, $\mu^{\text{RSO}}=2.6\Delta_0$,
          $\lambda=\Delta_0$, $L_{\text{RSO}}=0.5\xi_S$, $L_{\text{S}}=0.4\xi_S$, $h_l=0$, and $h_r=0.8\Delta_0$. }
	\label{fig1app}
	\end{figure*}
It also can transmit into the right F region
through processes marked by F-I i.e. particles and holes with up and
down spins. In the right region we set $h_r=0.8\Delta_0$ that causes
the energy splitting of spin-up and -down subbands. Let us first consider
the scattering process that occurs through $G^\downarrow_e$ channel
(the green curve in
Fig. \ref{fig1app}(b)). The density of states (DOS) in the left F region and the available
DOS of relevant subband in the right ferromagnetic
region determine the transmission probability through that specific
subband. The DOS within the left region in the absence of magnetization is
given by $N^{F_l}=2 |\varepsilon|/\pi\hbar^2v_F^2$ while in the right
region for the spin-down subband is $N_\downarrow^{F_r}=2 |\varepsilon-h_r|/\pi\hbar^2v_F^2$. We clearly
see that by increasing $eV$, $N^{F_r}$ linearly increases whereas $N_\downarrow^{F_r}$
decreases with an offset equal to $h_r$. The two DOSs are equal at
$h_r/2$ where the transmission through the spin-down subband is
maximal. From Fig. \ref{fig2app}(b) it is apparent that conductance through
$G^\downarrow_e$ channel reaches an extremum at
$eV=h_r/2=0.4\Delta_0$ which is fully consistent with the available
DOS discussion. Further increase in $eV$ causes the dominance
of $N_\downarrow^{F_r}$, which vanishes at $eV=h_r=0.8\Delta_0$ and results in
closing $G^\downarrow_e$ channel. When the voltage bias crosses one of
the Dirac cones belonging to each subband, since the DOS of this specific
subband vanishes at the Dirac point, the components of the conductance
feel this zero DOS. This vanishing DOS is seen as an abrupt
change in the components of conductance, depending on their own
available DOS, and thus the total conductance itself. 
	\begin{figure*}
	\includegraphics[width=12.30cm, height=7.0 cm
	]{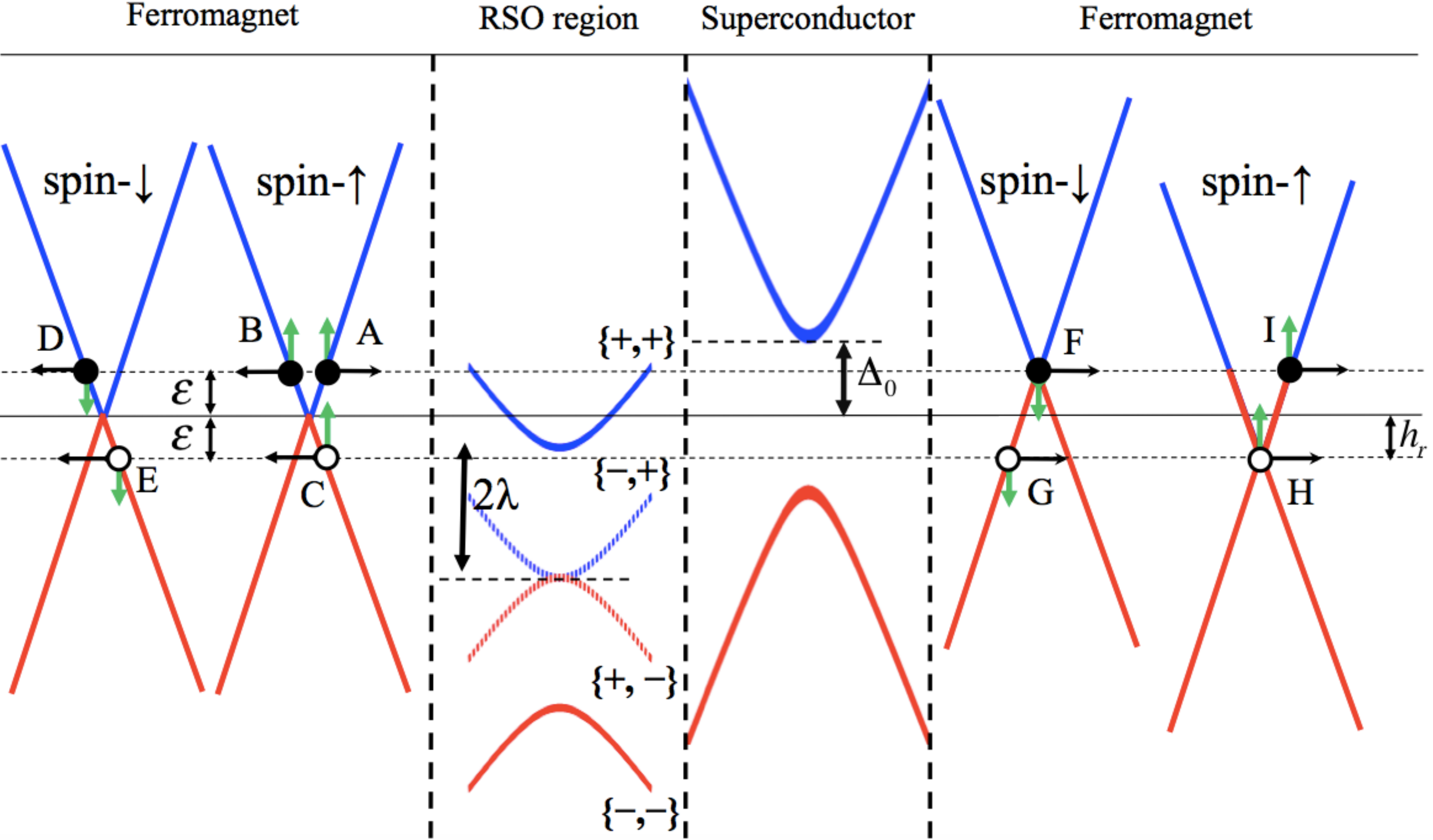}
	\caption{ Low energy band structure in each region. In the
          left region, spin-up and -down subbands are degenerated since
          $\mu^{F_l}=h_l=0$ while particles in the right F region belong to different subbands (spin-$\uparrow (\downarrow)$) so that the spin-down particles are minority spin spices (magnetization is orientated along the $z$ axis). In the RSOC region, there are two subbands splited by $2\lambda$, the strength of the RSOC, where the spin of excitations is locked to the direction of their momentum. In the superconductor segment, we consider a large doping so that the excitations above the superconducting gap $\Delta_0$ is described by a parabolic dispersion. The particle excitations are denoted by solid circles while the holes are shown by circles. Also their horizontal arrows represent propagation directions whereas the vertical arrows are their spin directions. The vertical axis is the energy of excitations ($\varepsilon$) while the horizontal one is their momentum (${\bf k}$). We introduce labels A-I to describe the scattering processes.}
	\label{fig2app}
	\end{figure*}
Within the
regime we considered, the
$N_\downarrow^{F_r}$ vanishes at $eV=h_r=0.8\Delta_0$ where we see that the
charge conductance Fig. \ref{fig2app}(a) and its
nonvanishing components Fig. \ref{fig2app}(b) show an abrupt change. We note that
similar analyses with identical conclusions can be made by
considering the critical values and limitations on the propagating
angles given by the transverse component of momentum
$q^c$ in the right and left F regions. It is worth mentioning that the
crossing of Dirac points, thus no (or less) available states for
quantum transport through the junction, and the appearance of abrupt changes in
conductance were also found in simple graphene-based
normal-superconductor junctions \cite{beenakkerrmp}  

Switching $h_l\neq 0$, one more abrupt change appears in the charge
conductance (not shown). This abrupt change can be also
understood by considering this fact that the magnetism
within the left region lifts the degeneracy of spin-up and -down subbands. Therefore, by varying the voltage bias, at $eV=\pm h_l$
the DOS of spin-down particles vanishes (the voltage bias reaches a
Dirac point). This analysis extends to
situations where $\mu^{F_r}$ and $\mu^{F_l}$ are nonzero. To perform
such analysis in the presence of nonzero chemical potentials, one needs
to carefully account for these quantities and determine how the
bandstructures 
change. We now proceed to analyze this more complicated case we employed
in the main text.    

	\begin{figure*}
	\includegraphics[width=12.30cm, height=7.0 cm
	]{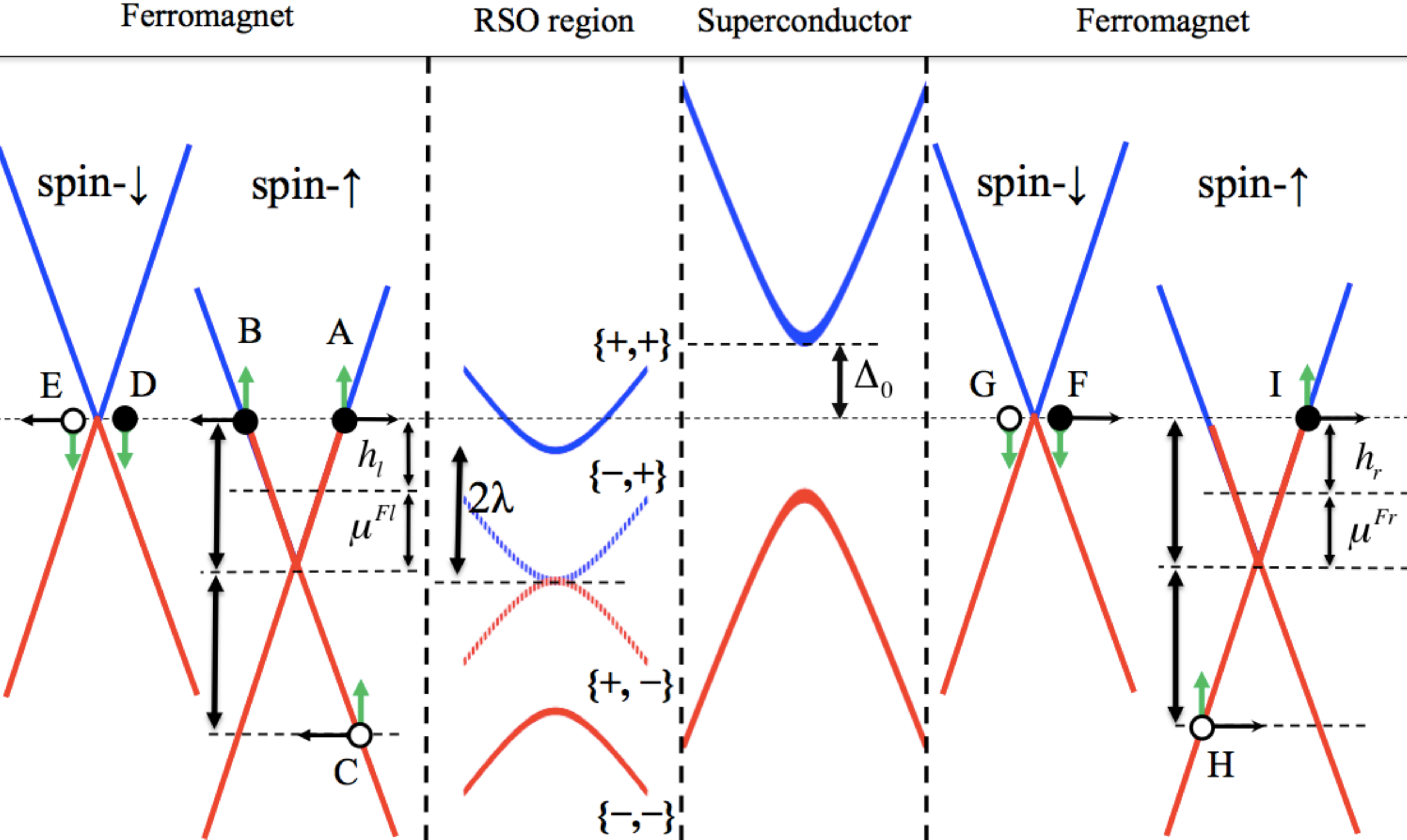}
	\caption{ Low energy band structure in each region. In the F regions, spin-up and -down particles belong to different subbands (spin-$\uparrow (\downarrow)$) so that the spin-down particles are minority spin spices (magnetization is orientated along the $z$ axis). In the RSOC region, there are two subbands splited by $2\lambda$, the strength of the RSOC, where the spin of excitations is locked to the direction of their momentum. In the superconductor segment, we consider a large doping so that the excitations above the superconducting gap $\Delta_0$ is described by a parabolic dispersion. The particle excitations are denoted by solid circles while the holes are shown by circles. Also their horizontal arrows represent propagation directions whereas the vertical arrows are their spin directions. The vertical axis is the energy of excitations ($\varepsilon$) while the horizontal one is their momentum (${\bf k}$). We introduce labels A-I to describe the scattering processes.}
	\label{fig3app}
	\end{figure*}

\*
\*
\\
Figure \ref{fig3app} illustrates the low-energy bandstructures in different regions of the
ferromagnet-RSOC-superconductor-ferromagnet junction with nonzero
$\mu^{F_l}, \mu^{F_r}, h_l,$ and $h_r$ similar to Fig. \ref{fig4} in the main
text. We here consider those parameters of Fig. \ref{fig1} in the main text so
that the anomalous CAR dominates i.e. $\varepsilon\ll \mu^{F_r}, h_r,
\Delta$ and $\mu^{F_r}=\mu^{F_l}=h_r=h_l$. In the F regions, the
magnetization splits the band structure into two subbands for
spin-up and -down excitations. We mark these subbands by
spin-$\uparrow\downarrow$. In the RSOC region, the spin and pseudo-spin
are coupled so that the spin-momentum locked bands are splitted and we
mark them by $\{\pm,\pm\}$ that refer to those $\pm$ appear in the
eigenvalues, Eq. (3) of the main text. The superconductor region is
assumed doped with a gap $\Delta_0$ in the energy spectrum. The same
as Fig. \ref{fig2app}, the solid circles represent particles while holes are shown
by circles. An incident particle with spin-up (A) can reflect back as
an electron with the same spin direction through process B with
amplitude $r_N^\uparrow$, with flipped spin direction $r_N^\downarrow$
(D), either as a hole with usual spin direction (E) with amplitude
$r_A^\downarrow$ or an anomalous Andreev reflection (C)
$r_A^\uparrow$. The incident spin-up particle (A) can also enter the
right ferromagnet as a spin-up electron $t_e^\uparrow$ (I), spin-down
electron $t_e^\downarrow$ (F), usual crossed Andreev reflection
$t_h^\downarrow$ (G), and anomalous crossed Andreev reflection
$t_h^\uparrow$ (H). As seen, because $\mu$ is set equal to $h$ in both
sides, the allowed transmission is $t_h^\uparrow$ which is consistent
with the propagation critical angles Eqs. (6) given in the main text and the
DOS analysis discussed above. 

The abrupt changes in the charge conductance occur when a hole passes
from the valance band into the conduction band. In other words, as discussed earlier, when the
voltage bias crosses a Dirac point, the corresponding DOS approaches
zero and therefore, an abrupt change appears. In the latter set of parameters we considered, one can introduce three regimes:
\begin{equation}
	\begin{cases}
	i, & 0\leq \varepsilon\leq|\mu^{F_r}-h_r| \\
	ii, & |\mu^{F_r}-h_r| \leq \varepsilon \leq |\mu^{F_r}+h_r| \\
	iii, & |\mu^{F_r}+h_r|\leq \varepsilon \\
	\end{cases}.
\end{equation}  
In regime $i$, an incident spin-up electron can be crossed Andreev
reflected with spin-up and -down in the conduction and valance bands,
respectively. By increasing the voltage bias $eV$, in regime $ii$,
spin-up crossed Andreev reflected hole remains in the conduction band
while the hole with spin-down moves into the valance band. In this
case, we expect an abrupt change in the charge conductance because
$eV$ crosses a Dirac point with zero DOS. It is worth mentioning that,
if the crossed Andreev reflected hole reisdes in the conduction band,
the reflection is of the retro type whereas if the crossed Andreev reflected hole
passes through the valance band, the reflection is of the specular
type. In other words, crossed Andreev reflected holes with different spins move
in different directions. In regime $iii$, both
crossed Andreev reflected holes with spin-up and -down
reside in the valance band. Therefore, one more abrupt change in the
charge conductance occurs when passing to this regime and a Dirac
point with zero DOS exists at $|\mu^{F_r}+h_r|$. Considering the
parameters of Fig. \ref{fig4}(a), $\mu^{F_r}=h_r$, the three regimes described
above reduce to two regimes that are separated at
$|\mu^{F_r}+h_r|=1.6\Delta_0$ where an abrupt change can be seen. Figure
\ref{fig4}(b) shows two abrupt changes because $|\mu^{F_r}-h_r|=0.4\Delta_0$ and
therefore a hole excitation experiences the three mentioned regimes
above. One abrupt change appears at $|\mu^{F_r}-h_r|=0.4\Delta_0$ and
the other at $|\mu^{F_r}+h_r|=1.2\Delta_0$.
\*
\*
\\
\end{widetext}
\*
\*
\\

\end{document}